\documentclass{elsart}

\usepackage{harvard}

\usepackage{graphicx}

\usepackage{amssymb}

\def\url#1{{\ttfamily\def\/{/\discretionary{}{}{}}#1}}

\def\aro{$\alpha_{RO} \; $}

\begin{document} 

\newcommand{\pn}{\par \noindent}

\def\ea{{\it et al.} }

\newcommand{\simgt}{\raisebox{-3.8pt}{$\;\stackrel{\textstyle > }{\sim }\;$}}
\newcommand{\simlt}{\raisebox{-3.8pt}{$\;\stackrel{\textstyle < }{\sim }\;$}}

\begin{frontmatter}
\title{On the Physical Conditions in\\ AGN Optical Jets}

\author[Scarpa]{Riccardo Scarpa\thanksref{rs}},
\author[Scarpa]{C. Megan Urry\thanksref{rs}}

\thanks[rs]{E-mail: scarpa@stsci.edu; cmu@stsci.edu}
\address[Scarpa]{Space Telescope Science Institute, 3700 San Martin Dr.,
Baltimore, MD 21218}

\begin{abstract}
The energy budget of all known optical jets is discussed. It is found 
that to power the extended radio lobes of radio galaxies, the jet bulk
motion on kpc scales must be relativistic, on average.  Based on
various constraints, a ``most probable'' region centered around
$\Gamma_{bulk} \sim 7.5$ and $\theta \sim 20^\circ$ is found. Because
of the consequent relativistic beaming, the rest 
frame magnetic field is lower and electron lifetimes longer.
Combining the effects of time dilation and lower emission rate, the
electron diffusion length becomes fully consistent with the
deprojected jet length, without the need for reacceleration.
\end{abstract}

\end{frontmatter}

\section{Introduction}
\label{intro}

To emit at optical-UV frequencies in a field of $B\sim
10^{-4}$ G, typical of the magnetic fields found in AGN jets on kpc scales,
electrons must have Lorentz factors $\gamma \simgt 10^6$ and
have diffusion times of a few hundred years.
Despite the shortness of the expected synchrotron cooling times, 
the jets are long and there is no indication of 
strong steepening of the radio-to-optical spectral
index as the distance from the nucleus increases 
(Sparks \ea 1996; Scarpa \ea 1999).
Therefore, either electrons are continuously reaccelerated
or the magnetic field is weaker than the equipartition value, as 
may be the case if relativistic beaming is important.\\
To shed light on this issue we analyze the energy
budget of all known optical jets, requiring they 
transport on average as much energy as needed to explain the
existence of extended radio lobes. Our approach relies on the
hypothesis that the extended radio structures are powered by jets
(Blandford \& K\"onigl 1979), and on the fact that, even if rare,
optical jets are discovered in all kinds of radio sources (of both FRI
and FRII morphologies and powers).

\begin{table*}
\tiny
\caption{Relevant Properties of Optical Jets$^{(a)}$}
\label{tab_data}
\begin{center}
\begin{tabular}{lrrrrrrrcrrrrr}
Name  & Cl.  & z  & $\nu_R$  & $F_R$  & $\nu_O$  & $F_O$  &
l & Vol  & $\alpha_{RO}$ & ref.  & $J$ & B$_{o}$  &B$_{r}$ \\
~ & (b)  &   & (c)  & (c)  & (d)  & (d)  &
(e) & (f)  & (g) & (h)  & (i) &  (j) & (k) \\
\hline 
M87         & I  &0.004 & 15  & 4.2  & 6.88 &1960 &  2.1 & 62.7&$ -0.71$& 2,24    &$ >450    $& 31  & 12 \\
3C~15       &I/II&0.073 & 2.78& 0.21 & 5.50 & 5.9 &  3.2 & 65.4&$ -0.86$& 21      &$ >50     $& 8.7 & 3.3\\
3C~66B      &I/II&0.021 & 15  & 0.04 & 8.17 & 6.5 &  3.0 & 65.1&$ -0.80$& 2,4,6,25&$ \simgt40$& 4.5 & 1.7\\
3C~78       & I  &0.029 & 1.66& 0.8  & 4.34 & 33  &  1.1 & 62.2&$ -0.81$& 12,26   &$ >40     $& 53  & 20 \\
3C~120      & I  &0.033 & 5.0 & 0.05 & 6.88 & 14  & 0.32 & 64.6&$ -0.69$& 11,13,27&$ >1000   $& 6.3 & 2.4\\
3C~200      & II &0.458 & 5.0 & 0.076& 4.16 & 18  & 0.32 & 65.9&$ -0.74$& 9,10    &$ >37.5   $& 15  & 5.8\\
3C~212      & II &1.049 & 8.33& 0.04 & 9.09 & 0.19& 0.85 & 67.1&$ -1.06$& 7       &$ \dots   $& 56  & 21 \\
3C~245      & II &1.029 & 5.0 & 0.045& 9.09 & 0.17& 31.7 & 67.2&$ -1.03$& 7,8,22  &$ \dots   $& 84  & 32 \\
3C~264      & I  &0.020 & 15  & 0.2  & 4.16 & 33  & 0.32 & 62.1&$ -0.85$& 1,2     &$ >37     $& 89  & 34 \\
3C~273      & II &0.158 & 15  & 2.41 & 5.50 & 57  &  4.8 & 64.8&$ -1.07$& 2,5,18  &$ >5300   $& 5.8 & 2.2\\
3C~346$^{(l)}$&II&0.161 & 15  & 0.08 & 4.16 & 11  & 11.0 & 62.8&$ -0.87$& 14,15   &$ \dots   $& 5.6 & 2.1\\
3C~371      & II &0.051 & 1.6 & 0.18 & 5.50 & 15  & 28.9 & 65.9&$ -0.74$& 17,19,23&$ >700    $& 17  & 6.4\\
0521--365   & I  &0.055 & 15  & 0.1  & 4.16 & 60  & 21.9 & 65.9&$ -0.73$& 3,16,23 &$ >10     $& 14  & 5.1\\
2201+044    & I  &0.027 & 5.0 & 0.009& 4.16 & 0.57& 2.05 & 63.9&$ -0.85$& 20,23   &$ \dots   $& 7.9 & 3.0\\
\hline
\end{tabular}
\end{center}
\scriptsize
{\bf (a)} All values are in the observer frame. {\bf (b)} Fanaroff \& Riley morphology class.
{\bf (c)} Radio frequency in GHz and radio flux in Jy.
{\bf (d)} Optical frequency in $10^{14}$~Hz and optical flux in $\mu$Jy. 
{\bf (e)} Projected jet length in kpc.
{\bf (f)} Logarithm of jet volume in cm$^3$, estimated 
    assuming cylindrical symmetry.
{\bf (g)} Radio to optical spectral index assuming 
    $F_\nu \propto \nu^\alpha$.\\
{\bf (h)} References for fluxes, size, and $J$: 1) Baum \ea 1997; 2) Crane \ea 1993; 
    3) Keel 1986; 4) Fraix-Burnet \ea 1989; 5) Bahcall \ea 1995; 
    6) Macchetto \ea 1991; 7) Ridgway \& Stockton 1997; 8) Laing 1989; 
    9) Burns \ea 1984; 10) Bogers \ea 1994; 11) Hjorth \ea 1995; 
    12) Sparks \ea 1995; 13) Fraix-Burnet \ea 1991; 
    14) van Breugel \ea 1992; 15) Dey \& van Breugel 1994; 
    16) Macchetto \ea 1991; 17) Wrobel \& Lind 1990; 18) Davis \ea 1991; 
    19) Akujor \ea 1994; 20) Laurent-Muehleisen \ea 1993; 
    21) Leahy \ea 1997; 22) Saikia \ea 1990; 23) Scarpa \ea 1999.; 
    24) Stiavelli \ea 1992; 25) Fraix-Burnet 1997;
    26) Saikia \ea 1986; 27) Walker \ea 1987.\\
{\bf (i)} Jet/counter-jet luminosity ratio as published or derived 
    by us from published radio maps.
{\bf (j)} ``Observed'' equipartition magnetic field in units of $10^{-5}$~G, 
    computed without transforming to the rest frame and without beaming.
{\bf (k)} ``Rest frame'' equipartition magnetic field in units of $10^{-5}$~G,
    computed assuming $\Gamma=7.5$ and $\theta = 20^\circ$ for all jets.
{\bf (l)} Only knot ``c'' as defined in van Breugel \ea (1992) considered.
\end{table*}
\normalsize

\section{Theoretical Considerations and Available Data}

Standard formulae for synchrotron emission are used (Pacholczyk 1970).  
The synchrotron emission is assumed to extend, in the observer
frame, from $\nu_1=10^7$ to $\nu_2=10^{15}$~Hz, following a single
power law ($F_\nu \propto \nu^\alpha$) of constant spectral index
$\alpha$.  The electron distribution in energy space
is $N(E)=N_0 E^p$, where $p=2\alpha - 1$.
Each electron emits in a narrow range of
frequencies centered on $\nu=c_1BE^2$, where $c_1=\frac{3e}{4\pi m^3c^5}$,
at a rate of $dE/dt=c_2B^2E^2$, where $c_2=\frac{2e^4}{3m^4c^7}$.
In the presence of relativistic beaming, rest frame ($L_R$)
and observed ($L$) jet luminosities are related by $L=L_R\delta^3$ (as
appropriate for a continuous jet; it would be $L=L_R\delta^4$ for a moving
sphere), where $\delta=[\Gamma(1-\beta \cos \theta)]^{-1}$ is the
Doppler beaming factor. The rest-frame luminosity is  
$L_R = \int_{E_1}^{E_2}{\frac{dE}{dt} N(E)dE}$,
where $E_1$ and $E_2$ are the cutoffs of the electron energy distribution 
corresponding to $\nu_1$ and $\nu_2$. After integrating, 
$L_R$ depends on $B$ and $N_0$, so that a second equation is 
necessary to solve completely for jet properties.
This can be obtained by imposing equipartition of energy, in 
which case we have $\frac{B^2\phi V}{8\pi} = (1+k)E_e$,
where $\phi$ is the magnetic field filling factor, V is the source volume,
and $E_e$ is the total electron energy. The proton energy, 
which remains unconstrained, is accounted for by 
the proton-to-electron energy ratio $k$.
For consistency with previous works, we set $k=0$ (assigning no energy 
to the protons).
Solving for $B$ and $N_0$ allows the calculation of the rest-frame
number density $n=\frac{L}{c_2 B^2 <E^2> V \delta^3}$ of the emitting 
electrons, which is
used to derive the kinetic energy transported in the jet, which is 
$L_k = \pi R^2 \Gamma^2 \beta c n <E>(1+k)$
(Celotti \& Fabian 1993). Here $\Gamma$ is the Lorentz factor
of the bulk motion, $\beta c$ the plasma speed, and $R$ the jet
radius. As before, the term $(1+k)$ accounts for the proton
energy. Under these quite standard assumptions,
the flux at 2 frequencies and the
bulk speed of the plasma suffice to evaluate the
jet kinetic energy.  Relevant data for all known optical jets are summarized
in Table 1.  Interestingly, optical jets are discovered in radio
sources of all types (Column 2), and have remarkably similar
radio-to-optical spectral indexes (\aro $\sim -0.8$), consistent with
radio observations where 90\% of the jets have $-0.5<\alpha <-0.9$ 
(Bridle \& Perley 1984). This indicates a high degree of
homogeneity of all jets, so that it is reasonable to compare
their energy budgets with those for radio lobes observed
in a large sample of radio sources.

\section{Bulk Jet Speed}

Relativistic beaming elegantly explains both jet one-sidedness and 
superluminal motion,
observed routinely on parsec scales (Zensus \& Pearson 1987; Readhead 1993),
and also on kpc scales in the jet of M87 (Biretta \ea 1999).
Assuming the two sides of the jet are intrinsically identical
(Rees 1978; Shklovsky 1970; Saslaw \& Whittle 1988; Laing 1988),
the Lorentz factor of the bulk speed can then be derived 
(Scheuer \& Readhead 1979) from the jet/counterjet luminosity ratio
$ J=\left(\frac{1+\beta \cos \theta}{1-\beta \cos \theta}\right)^{(2-\alpha)}$,
where $\theta$ is the angle between jet speed and line of sight, and
the exponent $(2-\alpha)$ is appropriate for a continuum jet (it would
be $3-\alpha$ for discrete emitting blobs; see Lind \& Blandford 1985).\\
The dependence of $J$ on jet inclination $\theta$ is
shown in Figure 1. It is seen that $J$ easily reaches very
large values even for modest bulk speed, but for large 
inclinations it remains finite and quite small even for $\Gamma
\rightarrow \infty$, so that $J$ can be effectively used
to constrain $\theta$. From Table 1 the median lower limit of $J$ 
for optical jets is 40, implying median inclinations  $\simlt 55^\circ$ 
and $\Gamma>1.1$.

\begin{figure}
\tiny
\includegraphics*[width=0.48\linewidth]{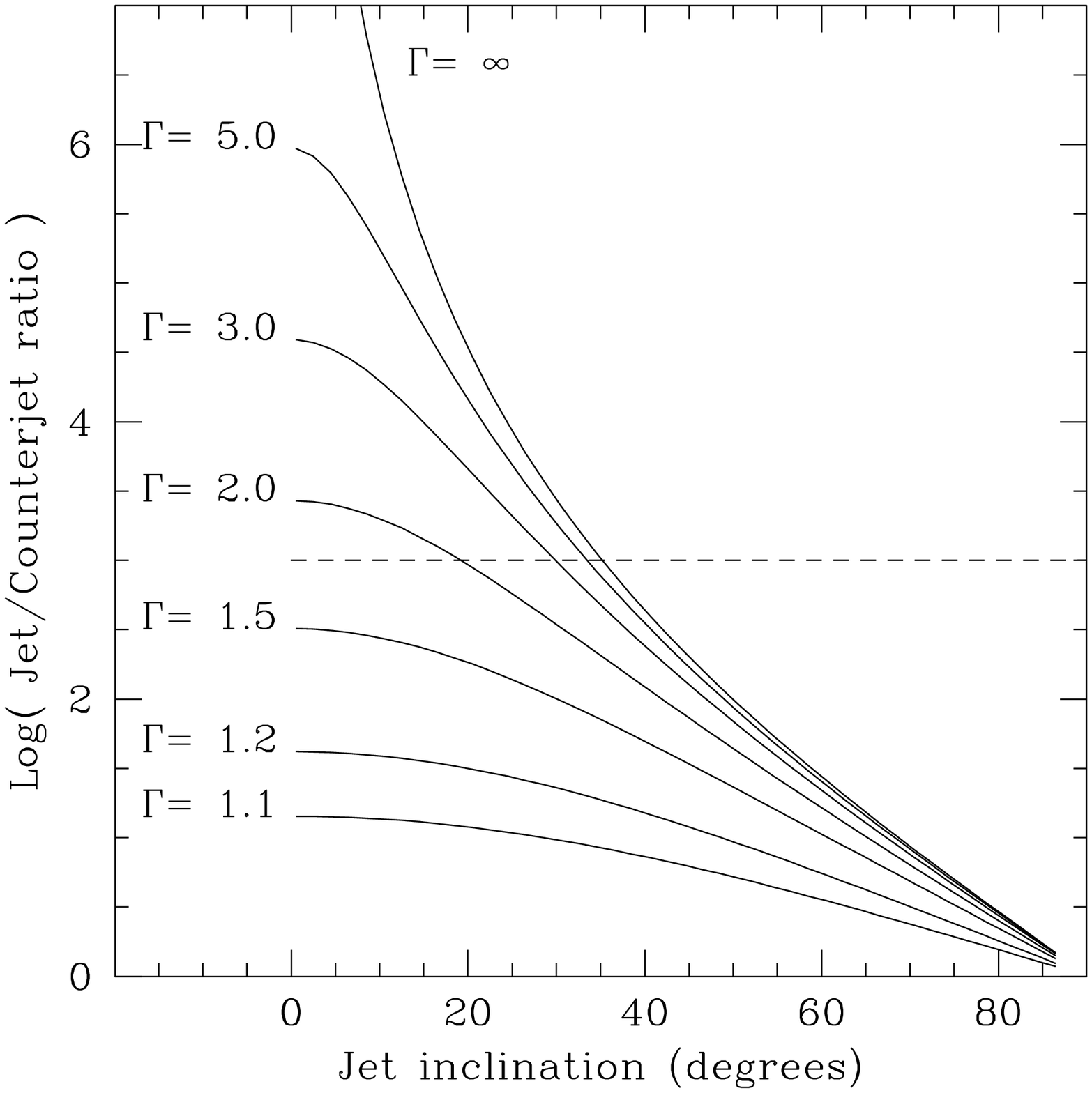}
\hspace{0.5cm}
\includegraphics*[width=0.49\linewidth]{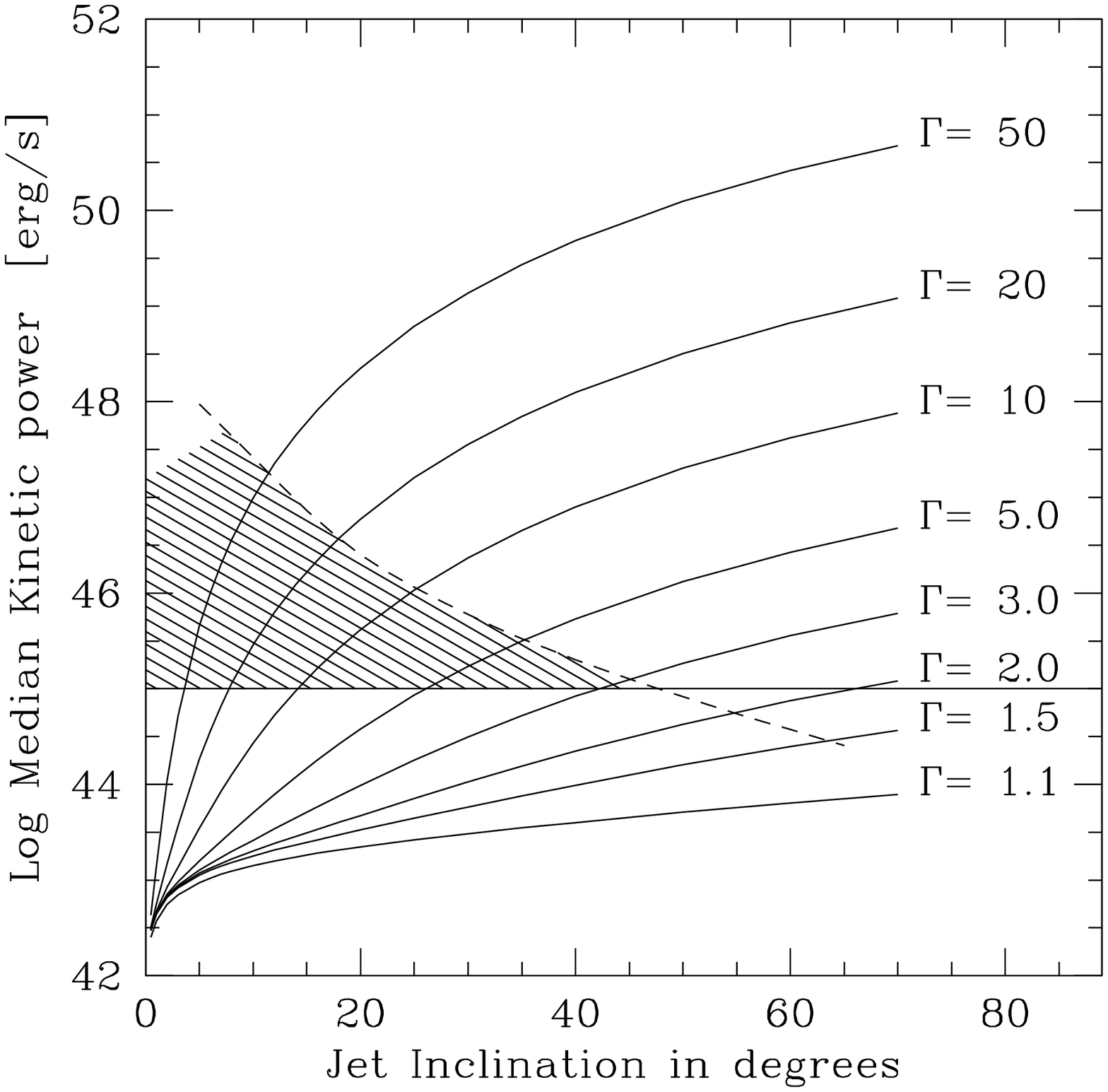}
\caption{
{\bf Left:} Jet/counterjet luminosity ratio $J$ for an intrinsically symmetric
jet, as function of jet inclination, for different Lorentz
factors for the bulk motion. Observed lower limits to $J$ give upper 
limits on $\theta$ and/or lower limits on $\Gamma$. For example the 
dashed line correspond to the available lower limits on $J$ for 3C~120.
This implies a jet inclination $<35^\circ$ and $\Gamma > 1.7$. 
{\bf Right:} Median kinetic power in
equipartition for the 14 optical jets versus jet inclination $\theta$
and for several values of $\Gamma$.
The horizontal line at $L_{kin}=10^{45}$ erg/s
indicates the median power estimated for radio lobes (Rawlings \&
Saunders 1991). The dashed line is the locus of beaming
$\delta = 1$, separating beamed (below) from de-beamed (above)
jets. Combining these two constraints, allowed $\theta$ and $\Gamma$
are restricted to the shaded area.
}
\normalsize
\end{figure}

\section{Jet Versus Lobe Kinetic Energy}

The kinetic luminosity depends on the density and average energy of
the emitting electrons, as well as the bulk speed of the plasma.
These quantities are therefore constrained if the kinetic
luminosity can be estimated in an independent way. We require
jets to transport enough energy for powering an average radio lobe.
Setting all relevant parameters as before
(i.e., same low frequency cutoff, $k=0$, and
$\phi=1$), a median kinetic power 
of $<L_{kin}>=10^{45}$~erg/s was found
for a large sample of radio galaxies including both
weak and powerful radio sources (Rawlings \& Saunders 1991). 
The high frequency cutoff
is higher for optical jets than radio lobes, 
but this should have no effect on the energy budget because for 
$p<-2$ ($\alpha<-0.5$) the energetics are 
dominated by the low energy particles.
Comparing the {\it median} kinetic power of all
optical jets with that of radio lobes (Figure 1), it is found 
that for very low bulk speeds ($\Gamma\simlt 2$),
the median kinetic energy for this sample of jets 
is at least one order of magnitude
smaller than needed to power average lobes, independently of
inclination angle. Very small inclinations are also
excluded, because for small $\theta$ the beaming
is so strong that the  rest frame luminosity
of the jet is very small, and the kinetic energy significantly reduced.\\
The maximum speed of the plasma can be loosely constrained if 
jets with enhanced rather than dimmed emission are preferentially discovered. 
For any given value of $\theta$, as soon as
$\Gamma$ becomes larger than the value defined by
$\Gamma-\sqrt{\Gamma^2-1} \cos \theta=1$, the emission is de-amplified, 
severely reducing our probability of discovering
the jet. In this way we derive a ``most probable'' region,
centered on $\theta \sim 20^\circ$ and $\Gamma \sim 7.5$,
where the average kinetic
energy carried by (optical) jets is fully consistent with the
requirement imposed by radio lobes. 
The conclusion is that jets should be 
relativistic at kpc scales.

\begin{figure}
\includegraphics*[width=0.5\linewidth]{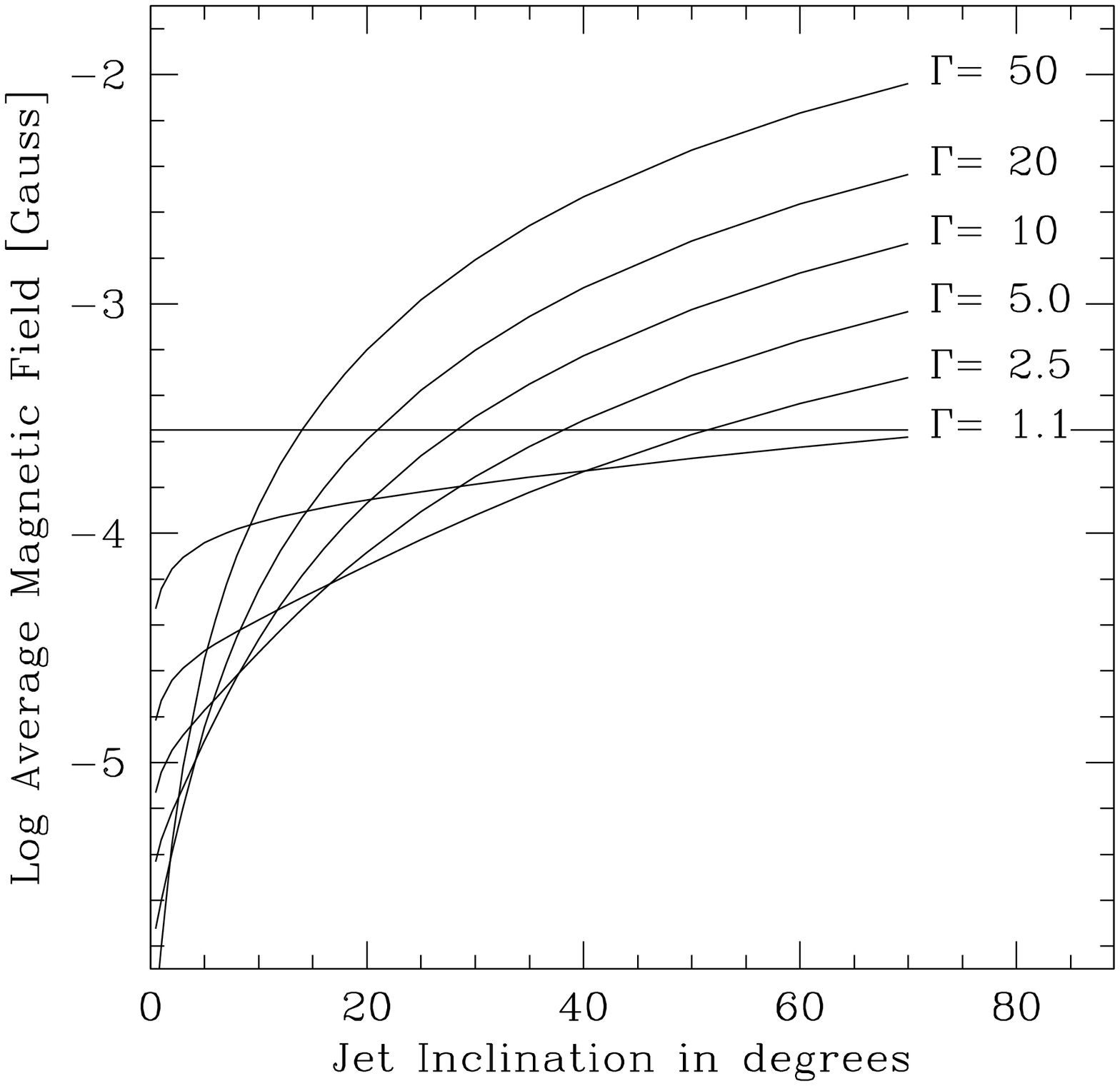}
\hspace{1cm}
\includegraphics*[width=0.245\linewidth]{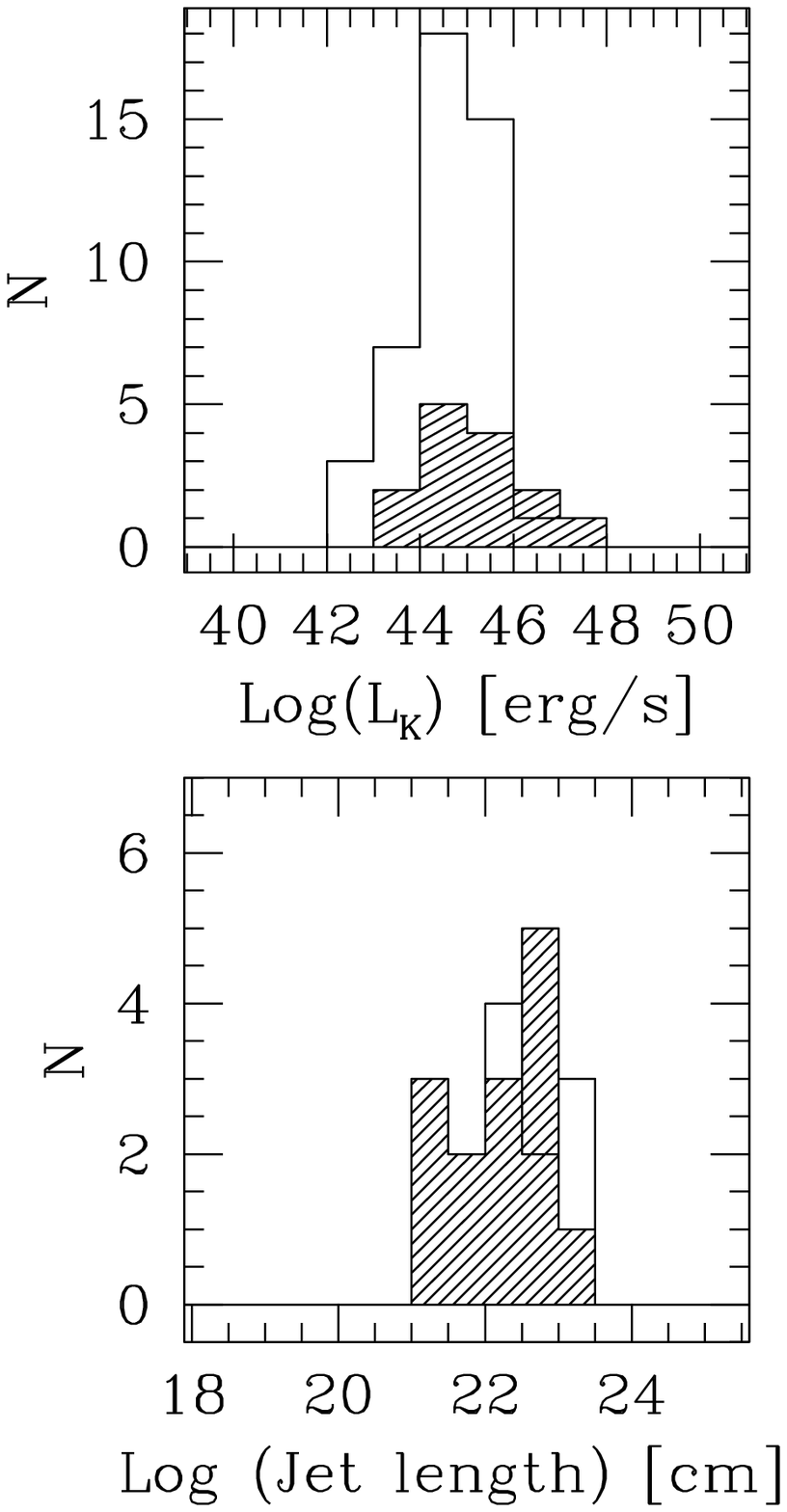}
\caption{ 
{\bf Left:} Average equipartition magnetic field for the whole sample 
versus jet inclination for several values of $\Gamma$.
The horizontal line at $B=2.8 \times10^{-4}$~G indicates the median
magnetic field derived directly from values given in Table 1 without
including beaming.
{\bf Upper right}: Kinetic energy distribution of jets (dashed histogram) 
compared to radio lobes power (open histogram), evaluated
assuming equipartition and $\theta=20^\circ$ and  $\Gamma=7.5$ for all jets.
The median jet kinetic energy is $10^{45}$ erg/s as 
for  radio lobes (Rawlings \& Saunders 1991).
{\bf Lower Right}: Distribution of estimated electron diffusion length 
$ct_{1/2}\Gamma$ for
electrons emitting at $10^{15}$ Hz in the observer frame (dashed histogram),
compared with the observed deprojected jet lengths (open histogram). 
The diffusion length was computed assuming  equipartition and
the same values of $\Gamma$ and $\theta$ as before.
}
\end{figure}

\section{Conclusions}

Comparing the kinetic power (as derived assuming equipartition) for
all known optical jets with that for typical radio lobes suggests
the presence of relativistic bulk motion of the emitting plasma at
$kiloparsec$ scales. In a non-relativistic scenario the estimated kinetic 
jet luminosity
is at least one order of magnitude less than needed to power the lobes.\\
This  has important consequences. 
Indeed,  at face value, the  data in Table 1 imply
strong magnetic fields and short electron lifetimes, leading to 
the conclusion that
electron reacceleration is necessary to explain the optical emission
on kpc scales (e.g., Meisenheimer \ea 1996). 
The analysis here points to an opposite possibility.
Indeed, the constraints on the kinetic luminosity allow
$\Gamma$ and $\theta$ to lie within a ``most probable'' region (Fig. 1),
centered near $\Gamma \sim 7.5$, corresponding to a highly relativistic 
bulk speed. Because of relativistic beaming, the rest-frame 
magnetic field is reduced (Table 1), and
the electron lifetimes are lengthened because of the lower energy losses
and time dilation. In these conditions the de-projected length $l/\sin
\theta$ of the jets is fully consistent with the electron diffusion length
$ct_{1/2}\Gamma$, without the need for reacceleration (Figure 2),
explaining the very nearly uniform $\alpha_{RO}$ observed
in the jets of M87 (Sparks \ea 1996) and PKS 0521-365 (Scarpa \ea 1999).

It is a pleasure to thanks G. Ghisellini, L. Maraschi, and F. Macchetto for 
helpful and encouraging comments. Support for this work was provided by NASA 
through grant  GO-06363.01-95A from the Space Telescope Science
Institute, which is operated by AURA, Inc., under NASA contract NAS~5-26555.

\end{document}